 \documentclass[aps,prl,twocolumn,groupedaddress]{revtex4-1}

\usepackage{lipsum}
\usepackage{graphicx} 
\usepackage{amsfonts}
\usepackage{amsmath}
\usepackage{braket}
\usepackage{gensymb}
\usepackage{hyperref}
\usepackage{multirow}

\begin{document}


\title{Observation of Discrete-Time-Crystal Signatures in an Ordered Dipolar Many-Body System}


\author{Jared Rovny}
\author{Robert L. Blum}
\author{Sean E. Barrett}
\email[]{sean.barrett@yale.edu}
\homepage[]{http://opnmr.physics.yale.edu/}
\thanks{}
\affiliation{Department of Physics, Yale University, New Haven, Connecticut 06511, USA}


\date{\today}

\begin{abstract}
A discrete time crystal (DTC) is a robust phase of driven systems that breaks the discrete time translation symmetry of the driving Hamiltonian. Recent experiments have observed DTC signatures in two distinct systems. Here we show nuclear magnetic resonance (NMR) observations of DTC signatures in a third, strikingly different system: an ordered spatial crystal. We use a novel DTC echo experiment to probe the coherence of the driven system. Finally, we show that interactions during the pulse of the DTC sequence contribute to the decay of the signal, complicating attempts to measure the intrinsic lifetime of the DTC.
\end{abstract}

\maketitle



Periodic driving of a many-body system may lead to interesting out-of-equilibrium states of matter. A recently described phenomenon that has received much attention is the discrete time-crystalline phase, in which a driven system spontaneously breaks the discrete time translation symmetry of its underlying Hamiltonian (also known as a Floquet time crystal, or a $\pi$-spin glass) \cite{Sacha2015, Khemani2016, vonKeyserlingk2016, Else2016, Moessner2017, Sacha2018}. Following a proposal for how to realize a discrete time crystal (DTC) \cite{Yao2017}, two experiments in quick succession showed evidence of DTC order. The first used trapped ions \cite{Zhang2017}, and closely followed original theoretical models for DTC behavior, specifically working in a regime thought to be conducive to many-body localization (MBL) \cite{Nandkishore2015}. The second used diamond nitrogen vacancy (NV) centers, which introduced several novel features: many spins, three-dimensional geometry, and long-range dipolar interactions \cite{Choi2017}. Although systems with these traits are considered unlikely candidates for MBL, sources of disorder did exist in that experiment, leaving some uncertainty as to the role of MBL in the observed signatures of DTC order \cite{Else2017, Ho2017, Huang2017, Russomanno2017}.

\begin{figure}
\includegraphics{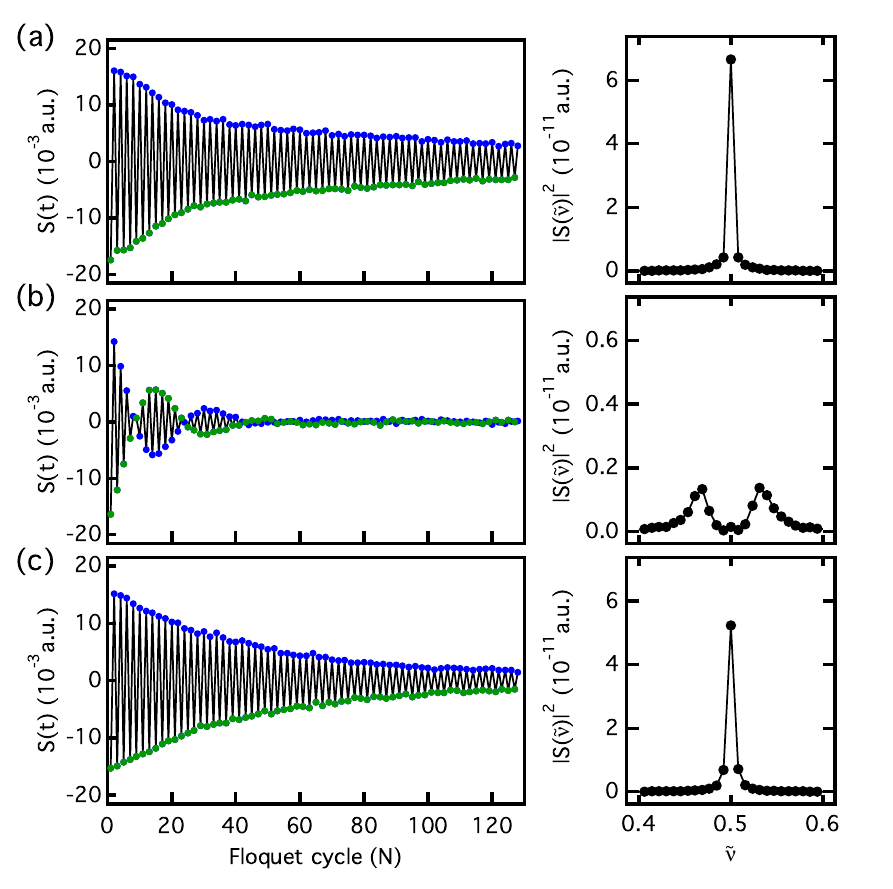}%
\caption{\label{CompareToLukin} After preparing the $^{31}$P spins in a $+z$ state, we measure the signal after the pulse sequence $\{\tau - X_\pi\}^N$. Green dots are for $N$ odd, blue are for $N$ even, and lines are to guide the eye. (a) For $\tau=12.5$ $\mu$s and $\theta = 0.995 \pi$, the response is an oscillation at half the drive frequency, corresponding to a single peak in the Fourier transform at normalized frequency $\tilde{\nu}=1/2$. (b) At the same $\tau=12.5$ $\mu$s but with $\theta = 1.054 \pi$, the signal exhibits a beat pattern, corresponding to a splitting of the Fourier peak. (c) For increased $\tau=392.5$ $\mu$s and $\theta=1.060\pi$, the sharp peak at $\tilde{\nu}=1/2$ is restored, despite the deviation of $\theta$ from $\pi$.}
\end{figure}

\begin{figure}
\includegraphics{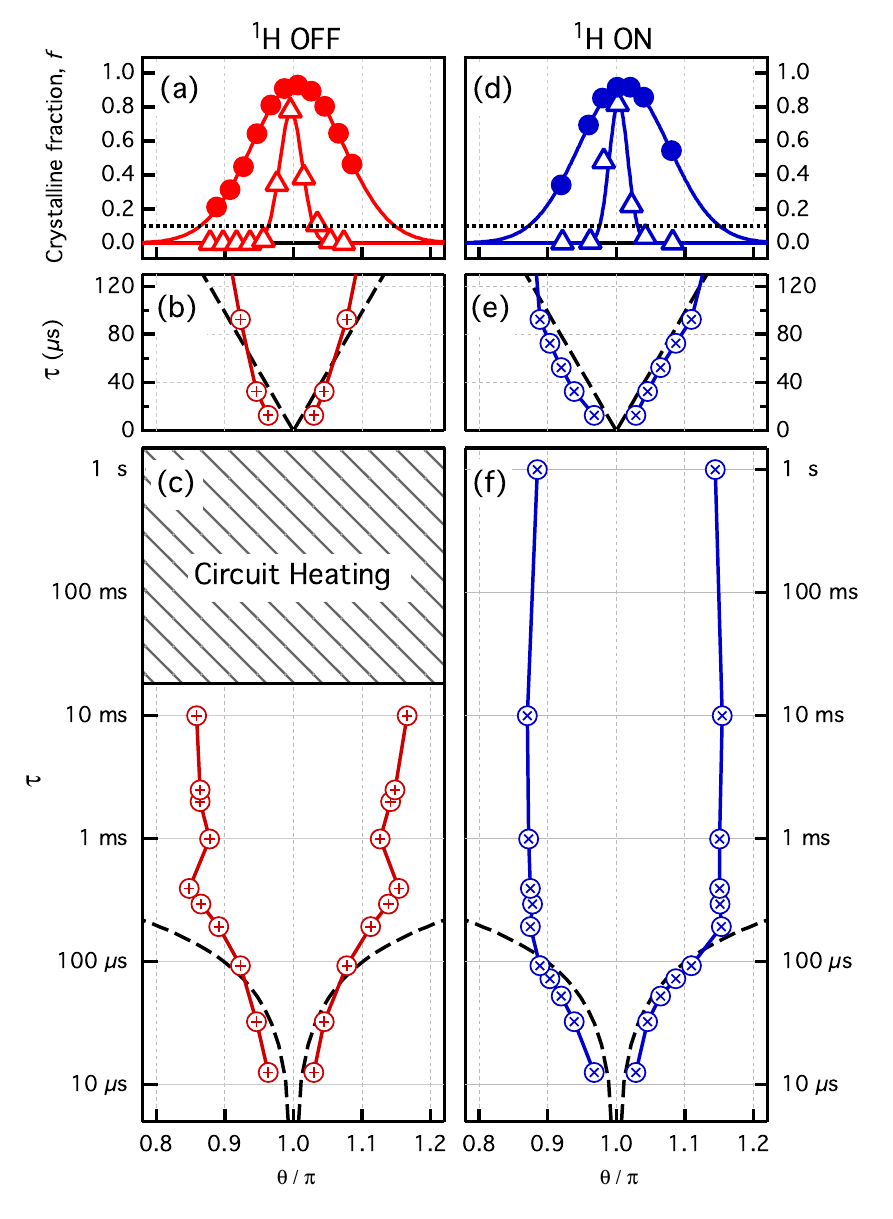}%
\caption{\label{PhaseDiagram} To explore the onset and robustness of oscillations at half the drive frequency, we examine the crystalline fraction $f$ using decoupling to turn the $^1$H off (left) and on (right). [(a) and (d)] The crystalline fraction varies smoothly as a function of $\theta / \pi$, fitting to Gaussians. Crystalline fractions from $\tau=12.5$ (open triangles) and $392.5$\,$\mu$s (closed circles) are shown. The horizontal dotted line shows $f=0.1$. [(b) and (c)] Cutoff at $f=0.1$ from the Gaussian fits, corresponding to the region in $\theta$ and $\tau$ within which we observe persistent DTC oscillations. With the $^1$H turned off, the boundary of this region shows structure near $\tau=1$ ms (see \cite{Rovny2018a} for further discussion). We also show $|\theta-\pi| =  W^\text{P,P} \tau$ [black dashes, (b) and (c), and (e) and (f)], where $W^\text{P,P}$ is the typical interaction scale of the $^{31}$P-$^{31}$P coupling (Table \ref{HamiltonianTable}). For $\tau > 10$ ms, the decoupling power begins to heat the tank circuit, skewing results, and preventing exploration of the $\tau > 10$\,ms region for $^1$H off. [(d)-(f)] Without decoupling, we can avoid heating and explore further in $\tau$ with $^1$H on. In (f), data span the range $0.03 < W^\text{P,P}\tau < 3200$\,radians.}
\end{figure}

In this Letter, we report the observation of discrete time-crystalline signatures in a system expected to be even further from the MBL limit than prior systems: an ordered spatial crystal. We observe robust oscillations at half the drive frequency (``DTC oscillations'' for brevity) across orders of magnitude in interaction time (Figs. \ref{CompareToLukin} and \ref{PhaseDiagram}). We also study the decay mechanism of the DTC oscillations, with two results. First, we show by generating a time-reversed DTC echo that the density matrix is more coherent than the original DTC sequence reveals. Second, we show that the effect of interactions during the nonzero pulse duration of the DTC sequence limits our ability to observe the intrinsic lifetime of the DTC oscillations.

We study the 100\% occupied crystal lattice of spin-1/2 $^{31}$P nuclei in ammonium dihydrogen phosphate (ADP), with chemical formula $\textrm{NH}_4\textrm{H}_2\textrm{PO}_4$. ADP is an ionic, tetragonal crystal that also contains spin-1/2 $^1$H nuclei (99.98\% abundant) and spin-1 $^{14}$N nuclei (99.64\% abundant) \footnote{We ignore the low natural abundance isotopes $^{15}$N (0.36\%), $^2$H (0.02\%), and $^{17}$O (0.04\%).}.

In the rotating frame \cite{Rovny2018a}, the secular internal spin Hamiltonian for $^{31}$P is
\begin{equation}\label{Hamiltonian}
  \mathcal{H}_\text{int} = \mathcal{H}_Z^\text{P} + \mathcal{H}_{zz}^{\text{P,P}} + \mathcal{H}_{zz}^{\text{P},\text{H}} + \mathcal{H}_{zz}^{\text{P},\text{N}}.
\end{equation}
Here, $\mathcal{H}_Z^\text{P}\approx \Omega I_{z_T}$ is a single-spin Zeeman term due to a uniform resonance offset \footnote{In general, the Zeeman shift can vary from site to site, but in our single-crystal system it is expected to be the same for all spins \cite{Rovny2018a, Eichele1994}}, and $\mathcal{H}_{zz}$ describes the various secular dipolar couplings between spins (where we omit terms that do not involve $^{31}$P) \cite{Slichter1996}. These dipolar coupling terms are summarized in Table \ref{HamiltonianTable}, with typical interaction frequencies $W/2\pi$ from the numerics that best match our experiments \cite{Rovny2018a}. Note that the dipolar coupling is a long-range interaction ($B_{ij}\propto r^{-3}$ in a three-dimensional system). During a strong rf pulse of phase $\phi$, an external pulse Hamiltonian $\mathcal{H}_{rf}=-\hbar \omega_1 I_{\phi_T}$ is added to Eq. (\ref{Hamiltonian}). Since $\omega_1 /2\pi \approx  68$ kHz is so large, the internal Hamiltonian terms are often ignored during the pulse action (the delta-function pulse approximation) \cite{Slichter1996, Li2007, Li2008}.  We revisit this approximation later.

\begin{table}
  \caption{\label{HamiltonianTable}Terms in the Hamiltonian involving $^{31}$P nuclear spins in ADP. The dipolar coupling constant for a pair of spins is
    $B_{ij}=\frac{\mu_0}{4\pi}\frac{\gamma_i\gamma_j \hbar^2}{|\vec{r}_{ij}|^3}\frac{1}{2}[1-3 \textrm{cos}^2(\theta_{ij})]$,
  where $\theta_{ij}$ is the angle between the internuclear vector $\vec{r}_{ij}$ and the $z$ axis (defined by the static external field), $\mu_0$ is the vacuum permeability, and $\gamma_i$ and $\gamma_j$ are the nuclear gyromagnetic ratios of the two spins. $\{I_{\phi}, S_{\phi}, R_{\phi}\}$, $\phi=x,y,z$, are the spin operators for $\{^{31}$P$, ^1$H$,^{14}$N$\}$. In the presence of the external $H_0=4$\,T field, the Larmor frequencies $\omega_{0}/2\pi=\gamma H_0/2\pi$ of $^{14}$N, $^{31}$P, and $^{1}$H nuclei are 12, 69, and 170 MHz respectively.}
\begin{ruledtabular}
  \begin{tabular}{l r r}
    Hamiltonian term  & \multicolumn{2}{r}{Typical interaction freq. (Hz)}\\
  \hline
  \multicolumn{2}{l}{$\mathcal{H}_{zz}^{\text{P},\text{P}} = \sum_{i,j>i}B^\text{P}_{ij} ( 3I_{z_i}I_{z_j} - \vec{I}_i\cdot\vec{I}_j )$}  & $W^\text{P,P}/2\pi=508$ \\
  \multicolumn{2}{l}{$\mathcal{H}_{zz}^{\text{P},\text{H}} = \sum_{i,j}B^\text{H}_{ij} (2I_{z_i}S_{z_j})$} & $W^\text{P,H}/2\pi=3500$ \\
  \multicolumn{2}{l}{$\mathcal{H}_{zz}^{\text{P},\text{N}} = \sum_{i,j}B^\text{N}_{ij} (2I_{z_i}R_{z_j})$} & $W^\text{P,N}/2\pi=97$
  \end{tabular}
\end{ruledtabular}
\end{table}

There are four primary differences between the spin Hamiltonian in this system and prior models. First, the spin Hamiltonian is exceptionally ordered. The Zeeman term is extremely uniform in space, and the symmetry of the ADP crystal causes each $^{31}$P nucleus to experience an identical set of dipolar couplings from other $^{31}$P, $^{1}$H, and $^{14}$N \cite{Rovny2018a}. Second, rather than having only an Ising-type term $I_{z_i}I_{z_j}$, the dipolar coupling between the $^{31}$P nuclei includes ``flip-flop'' terms $I_{x_i}I_{x_j}+I_{y_i}I_{y_j}=(I^+_{i}I^-_{j}+I^-_{i}I^+_{j})/2$. We exploit this difference to make echoes, as we describe below. Third, while the detected spin-1/2 $^{31}$P spins interact among themselves, they also interact with two other spin species that are not directly affected by the repeated pulses: the spin-1 $^{14}$N (with a weaker coupling) and the spin-1/2 $^1$H (with a stronger coupling). Fourth, we are able to selectively use high power continuous wave (cw) decoupling to ``effectively remove'' the $^1$H from the spin Hamiltonian; we label this ``$^1$H off.'' Experiments where this technique is not used are labeled ``$^1$H on.'' 

To do these experiments, we first prepare the $^{31}$P spins in a weakly polarized $+z$ state (not a pure state) by cross polarizing the $^{31}$P spins with the bath of room-temperature $^1$H spins, producing an initial $+y$ polarization of $^{31}$P, then applying a final pulse to rotate the magnetization into $\hat{z}$ \cite{Rovny2018a}. While this procedure improves the initial polarization by a modest factor $\gamma_\textrm{H}/\gamma_\textrm{P}\approx2.5$, it significantly increases our ability to quickly repeat experiments, as the relaxation time $T^\text{H}_{1} = 0.6$\,s of the $^1$H is much faster than that of the phosphorus, $T^\text{P}_1 = 103$\,s.

Beginning with the enhanced $^{31}$P spin polarization, we apply the pulse sequence $\{\tau-X_\theta\}^N$, where $\tau$ is a wait time, and the pulse $X_\theta$ is a rotation about $+\hat{x}$ by angle $\theta$. After $N$ such repeated Floquet cycles, we apply a final $X_{\pi/2}$ to generate a nuclear magnetic resonance (NMR) signal that reveals the $z$ magnetization after $N$ cycles. Unless otherwise stated, we apply $^1$H decoupling \footnote{For cw decoupling, we use $\gamma_\textrm{H} H_1 / 2 \pi \approx 18$\,kHz.} from the end of cross polarization through the readout. After reinitializing the starting state over time $\approx 5T_1^\text{H}$, the pulse sequence is repeated for the next $N$, from $N=1$ to $N=128$, producing a discrete time signal $S(t)$ with Fourier transform $S(\nu)$. In practice, the internal Hamiltonian [Eq. (\ref{Hamiltonian})] acts not only during wait time $\tau$, but also during the pulse, which is of duration $t_p\approx 7.5$\,$\mu$s. The total Floquet period is $T=\tau + t_p$, with drive frequency $\nu_F = 1/T$; below, we examine the spectrum $S(\tilde{\nu})$ as a function of the normalized frequency $\tilde{\nu}\equiv \nu / \nu_F$.

At small $\tau$ and $\theta \approx \pi$, the system responds with trivial oscillations at half the drive frequency, i.e., at $\tilde{\nu}=1/2$ [Fig. \ref{CompareToLukin}(a)]. As $\theta$ is adjusted away from $\pi$, the response beats with frequency determined by $\theta - \pi$ [Fig. \ref{CompareToLukin}(b)], also an expected result. However, with $\theta \neq \pi$ and increased values of $\tau$, the system returns to robust $\tilde{\nu}=1/2$ oscillations, a signature of DTC behavior \cite{Yao2017} [Fig. \ref{CompareToLukin}(c)].

We explore this behavior as a function of $\tau$ and $\theta$ by examining the ``crystalline fraction'' $f=|S(\tilde{\nu} = 1/2)|^2/\sum_{\tilde{\nu}}|S(\tilde{\nu})|^2$, as defined in \cite{Choi2017}, except that we Fourier transform our entire $S(t)$ rather than just a late-time window (see \cite{Rovny2018a} for further discussion of this point). For given $\tau$, we fit the crystalline fraction as a function of $\theta$ to Gaussians with good results [Fig. \ref{PhaseDiagram}(a)]. Note that the precise shape of $f(\theta)$ does depend on the choice of Fourier transform window size, as described in \cite{Rovny2018a}. These Gaussians reveal a region within which the robust $\tilde{\nu}=1/2$ oscillations are detectable (the ``DTC region''), and outside of which there are diminished or split Fourier peaks. We may visualize the DTC region [shown in Figs. \ref{PhaseDiagram}(b) and \ref{PhaseDiagram}(c)] by choosing a cutoff $f=0.1$. The width of the DTC region increases with $\tau$ at short $\tau$, fluctuates slightly near $\tau=1$\,ms (discussed further in \cite{Rovny2018a}), and then approaches a steady value at long $\tau$.

Although we explore multiple decades with $^1$H off, heating of the NMR tank circuit (from cw decoupling) imposes an experimental barrier for $\tau > 10$\,ms \cite{Rovny2018a}. In order to explore further in $\tau$, we turn off the decoupling, allowing the $^1$H to act. We prepare the initial state in the same way as described above, but then apply pulses to the $^{31}$P only [see Figs. \ref{PhaseDiagram}(d)-\ref{PhaseDiagram}(f)]. In this $^1$H-on case, the DTC region approaches its maximum width faster as a function of $\tau$, and shows little to no fluctuation in width over many decades in $\tau$. The width appears to decrease slightly when the experiment time approaches $T_1^\textrm{P}$, the relaxation time of the $^{31}$P spins.

\begin{figure}
\includegraphics{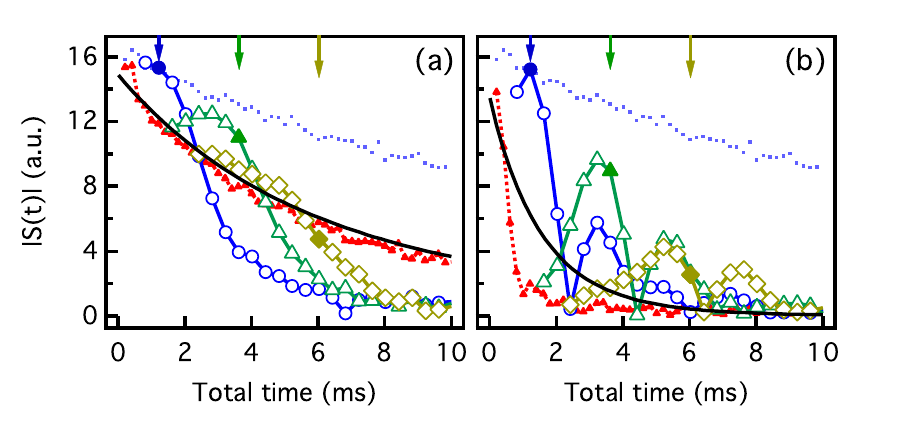}%
\caption{\label{Echoes} A decay envelope $\textrm{cos}^N(\epsilon)$ (solid black line) bounds the magnitude of our DTC oscillations (closed red triangles) for long $\tau$ and large $\epsilon$. To show that the observed decay is not dominated by an irreversible process, we devise an approximate ``unwrapping'' sequence to create an echo above the classical decay envelope. After $N$ cycles of $\{\tau - X_\theta\}$ (closed red triangles), we switch to the second part of the echo sequence (see text) and increment $N'$. The results of this $N'$ sequence are shown for $\tau = 192.5$ $\mu$s, where clear echoes are observable for both $\theta = 1.08\pi$ (a) and $\theta=1.16\pi$ (b), with $N=$ \{2, 6, 10\} (open blue circles, green triangles, and yellow diamonds). The corresponding arrow and filled marker show the predicted time of the DTC echo maximum, i.e., where $N'=N$. For the case of the DTC sequence at $\theta=\pi$ (blue dots), we observe decay where none is expected in the perfect, delta-function pulse model; the echoes do not rise above this $\theta=\pi$ decay envelope. We examine this effect in Fig. \ref{XYpulse}.}
\end{figure}

While these robust oscillations for many $\tau$ are distinctive signatures of a discrete-time-crystalline phase, the lifetime of the state is of particular interest. To explore reasons for the decay of oscillations during the DTC experiment, we begin by considering a trivial decay mechanism for noninteracting spins \cite{Rovny2018a}. If we start with a net magnetization vector aligned along $\hat{z}$, then an $X_{\pi+\epsilon}$ pulse leaves a component of the magnetization along the $z$ axis proportional to $\textrm{cos}(\epsilon)$. If the component of the magnetization in the transverse plane after the pulse is assumed to dephase and be lost during the subsequent $\tau$ due to some variation in local fields, we expect each Floquet cycle to reduce the observable magnetization by $\textrm{cos}(\epsilon)$, leading to an exponential decay $\textrm{cos}^N(\epsilon)$ after $N$ Floquet cycles. In practice, our signals seem to stay at or below the barrier imposed by this decay rate (compare Fig. \ref{Echoes}, closed red triangles versus black solid line).

To test whether magnetization is irreversibly lost during the decay of the DTC oscillations, we devise a ``DTC echo'' sequence that attempts to ``undo'' the forward evolution of the system during the original DTC sequence. In the original sequence $\{\tau-X_\theta\}^N$, the system evolves under $\mathcal{H}_\text{int}$ for time $\tau$, followed by a pulse rotation. To undo the rotation, we simply apply a pulse rotation in the opposite sense, $\overline{X}_\theta$. To undo the effect of $\mathcal{H}_\text{int}$ during $\tau$, we first limit our attention to the homonuclear interaction $\mathcal{H}^{\textrm{P},\textrm{P}}_{zz}$, and we borrow a technique from the so-called ``magic echo'' experiment in NMR \cite{Slichter1996, Rhim1971}, exploiting the fact that a strong rf pulse along $y$ reduces $\mathcal{H}_{zz}^\text{P,P}$ to an effective internal Hamiltonian term $-\frac{1}{2}\mathcal{H}^{\textrm{P},\textrm{P}}_{yy}$, where we define $\mathcal{H}^{\textrm{P},\textrm{P}}_{\phi \phi}=\sum_{j>i}^NB^\textrm{P}_{ij} ( 3I_{\phi_i}I_{\phi_j} - \vec{I}_i\cdot\vec{I}_j )$ \cite{Rovny2018a}. By allowing $-\frac{1}{2}\mathcal{H}^{\textrm{P},\textrm{P}}_{yy}$ to act for time $2\tau$ and rotating the state with surrounding $X_{\pi/2}$ pulses, the net phase evolution becomes $-\mathcal{H}^{\textrm{P},\textrm{P}}_{zz}\tau$, opposite the original forward dipolar evolution $+\mathcal{H}^{\textrm{P},\textrm{P}}_{zz}\tau$. The final DTC echo sequence is
\begin{equation}
\{\tau - X_\theta\}^N- (X_{\pi/2} - \{ \overline{X}_\theta - Y_{\Phi} \}^{N'} - \overline{X}_{\pi/2}),
\end{equation}

where $\Phi=\omega_12\tau$ for a strong rf pulse of duration $2\tau$. In the language of the more conventional Hahn spin echo sequence, the first part ($N$ blocks) of this sequence generates the ``free induction decay,'' while the second part (the rotated $N'$ blocks) generates the signal ``after the $\pi$ pulse'' \cite{Hahn1950,Slichter1996}. If the second part of this sequence can undo the first part, then we expect an echo to appear when $N'=N$. In Fig. \ref{Echoes}, we show prominent echoes in the magnetization for various values of $N$. The detection of these DTC echoes is evidence that the original DTC sequence drives coherence to normally unobservable parts of the density matrix, without being irreversibly lost due to interactions with an external thermal bath.

Note that even at $\theta = \pi$, the signal decays (Fig. \ref{Echoes}), which should not happen in the delta-function pulse approximation. Moreover, this $\theta=\pi$ decay envelope appears to limit the recovery of signal for $\theta \neq \pi$, where we see that the DTC echoes are unable to cross this barrier (Fig. \ref{Echoes}). This may be also responsible for the echo peaks occurring earlier than their expected locations. An obvious mechanism that could cause this is pulse imperfections; however, our studies have shown their effects to be too small to explain the observed decay \cite{Rovny2018a}.

\begin{figure}
\includegraphics{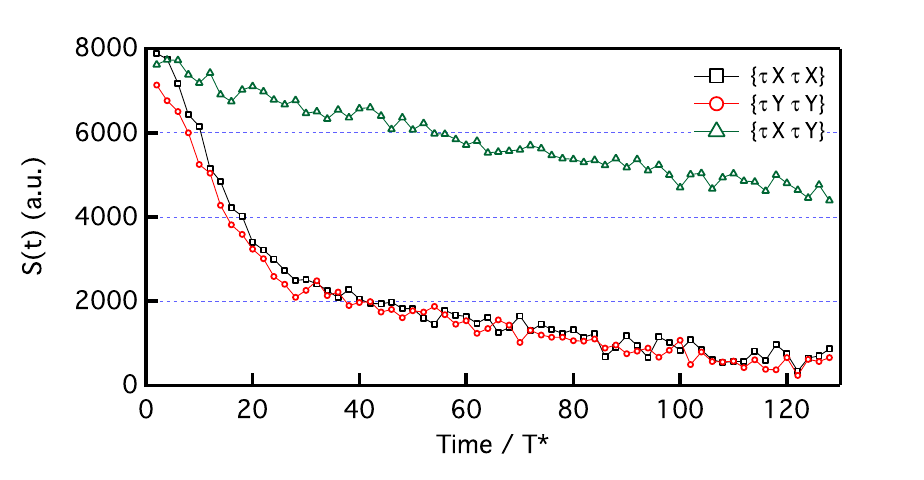}%
\caption{\label{XYpulse} Experimental pulses are of nonzero duration, which allows the internal Hamiltonian to act. These actions, while small, produce significant effects, especially after many Floquet cycles. At $\tau=20$ $\mu$s, pulse sequences $\{X,X\}$ (black open squares) and $\{Y,Y\}$ (red open circles) decay much faster than the sequence $\{X,Y\}$ (green open triangles). Signals are sampled after each repeating block $\{\alpha, \beta \}$. Here, our $T^* = \tau + t_p$.}
\end{figure}

In order to account for this decay, we reexamine the previously ignored effects of $\mathcal{H}_\text{int}$ during the nonzero duration pulses. Experimentally, we explore this by using modified versions of the DTC sequence, using different sets of pulse phases at $\theta = \pi$. Defining $\{\alpha,\beta\}\equiv\{\tau - \alpha_\pi - \tau - \beta_\pi\}^N$, we expect little or no difference between $\{X,X\}$, $\{Y,Y\}$, and $\{X,Y\}$ for ideal delta-function pulses. However, Fig. \ref{XYpulse}(a) shows that while $\{X,X\}$ and $\{Y,Y\}$ produce very similar results, $\{X,Y\}$ yields a dramatically extended lifetime.

This result can be qualitatively understood when accounting for the nonzero pulse duration $t_p$. During a pulse of phase $X$ ($Y$), an effective internal Hamiltonian $-\frac{1}{2}\mathcal{H}_{xx}^{\text{P,P}}$ ($-\frac{1}{2}\mathcal{H}_{yy}^{\text{P,P}}$) acts for time $t_p$ \cite{Li2007, Li2008}. For $\{X,X\}$, a term proportional to $-\frac{1}{2} \mathcal{H}_{xx}^{\text{P,P}}$ is present in the average Hamiltonian; this term is $-\frac{1}{2} \mathcal{H}_{yy}^{\text{P,P}}$ for $\{Y,Y\}$. However, for $\{X,Y\}$ this term is instead proportional to $ -\frac{1}{2}(\mathcal{H}_{xx}^{\text{P,P}}+\mathcal{H}_{yy}^{\text{P,P}})= +\frac{1}{2}\mathcal{H}_{zz}^{\text{P,P}}$ \cite{Slichter1996}, which commutes with the initial density matrix. We explore the effect of nonzero pulse duration further in \cite{Rovny2018a}.

The results we have shown are strikingly similar to those described by Zhang \textit{et al}.\ \cite{Zhang2017} and Choi \textit{et al}.\ \cite{Choi2017}, even though the spin Hamiltonian for our system is different in interesting ways. We also explore a very large region in the $(\theta,\tau)$ parameter space, and observe DTC signatures across a remarkably broad range in $\tau$. This occurs despite the conventional wisdom that our experimental conditions should be even less conducive to MBL than all prior DTC experiments \cite{Ho2017,Nandkishore2015}. DTC oscillations like those observed in this experiment can sometimes be explained as prethermal phenomena, but our system seems to violate conditions for prethermalization \cite{Else2017,Rovny2018a}. The DTC echo opens a new window into the physics of these driven systems, and helps to clarify the nature of the DTC oscillations. Further, we have shown that the interactions during the pulse of the DTC sequence contribute to the decay of the oscillating signal. This is a practical barrier to measuring the intrinsic lifetime of the DTC, which should be taken into account in future studies.

In this work, we exploited both the long coherence times of our sample, and our ability to use NMR pulse sequences to edit the spin Hamiltonian.  This suggests that NMR can be a useful probe of the physics of out-of-equilibrium, driven many-body systems. 

\begin{acknowledgments}
  We thank C. W. von Keyserlingk, V. Khemani, C. Nayak, N. Yao, and M. Cheng for helpful discussions. We also thank C. Grant and D. Johnson for help in constructing the NMR probe, K. Zilm for recommending the ADP sample, and S. Elrington for assistance with implementing cross polarization. This material is based upon work supported by the National Science Foundation under Grant No. DMR-1610313. R.L.B. acknowledges support from the National Science Foundation Graduate Research Fellowship under Grant No. DGE-1122492. 
\end{acknowledgments}

\textit{Note added}.---Recently, the authors of an interesting related experiment contacted us, alerting us to their liquid state NMR search for temporal order of periodically driven spins in star-shaped clusters \cite{Pal2018}. They study a unique spin Hamiltonian, and they explore a range of cluster sizes (with $N=1,4,10$, and $37$ spins).


%

\end{document}